\documentclass[aps,twocolumn,amsmath,amssymb,floatfix,pra,reprint,showpacs,footinbib,superscriptaddress,longbibliography]{revtex4-1}
\usepackage[usenames,dvipsnames]{color}
\usepackage{psfrag,graphicx}
\usepackage{dcolumn}
\usepackage{bm}
\usepackage{amsfonts,amssymb,amsmath}
\usepackage{slashed}
\usepackage{mathtools}
\usepackage[colorlinks]{hyperref}
\usepackage{pgfplots}
\hypersetup{%
        plainpages=true,
        breaklinks=true,% not default in dvips mode, so we must specify
        hypertexnames=false,%not ideal, but needed when pagenums duplicate (`i' vs. `1')
        pageanchor=true,
        colorlinks=true,
        linkcolor={blue},
        citecolor={red},
        urlcolor={blue},
        pagecolor={black},
        anchorcolor={black}
      }

\newcommand{\be}{\begin{equation}}
\newcommand{\ee}{\end{equation}}
\newcommand{\bq}{\begin{eqnarray}}
\newcommand{\eq}{\end{eqnarray}}

\newcommand{\ket}[1]{\left |#1 \right\rangle}
\newcommand{\bra}[1]{\left \langle #1 \right |}

\newcommand{\ketbra}[2]{|#1\rangle\langle#2| }

\newcommand{\equ}[1]{Eq.~(\ref{#1})}

\newcommand{\beq}{\begin{eqnarray}}
\newcommand{\eeq}{\end{eqnarray}}
\renewcommand{\eqref}[1]{\mbox{Eq.~(\ref{#1})}}

\begin{document}
\title{Amplified opto-mechanical transduction of virtual radiation pressure}

\author{Mauro Cirio}
\affiliation{CEMS, RIKEN, Wako-shi, Saitama 351-0198, Japan}

\author{Kamanasish Debnath}
\affiliation{CEMS, RIKEN, Wako-shi, Saitama 351-0198, Japan}
\affiliation{Amity Institute of Applied Sciences, Amity University, Noida - 201303 (U.P.), India}
\affiliation{Laboratory of Theoretical Physics of Nanosystems, Ecole Polytechnique F\'ed\'erale de Lausanne (EPFL), CH-1015 Lausanne, Switzerland}

\author{Neill Lambert}
\email[e-mail:]{nwlambert@gmail.com}
\affiliation{CEMS, RIKEN, Wako-shi, Saitama 351-0198, Japan}

\author{Franco Nori}
\affiliation{CEMS, RIKEN, Wako-shi, Saitama 351-0198, Japan}
\affiliation{Department of Physics, University of Michigan, Ann Arbor, Michigan 48109-1040, USA}

\date{\today}
\begin{abstract}
Here we describe how, utilizing a time-dependent opto-mechanical interaction, a mechanical probe can provide an amplified measurement of the virtual photons dressing the quantum ground state of an ultra strongly-coupled light-matter system. We calculate the thermal noise tolerated by this measurement scheme, and discuss a range of experimental setups in which it could be realized.
\end{abstract}
\maketitle

{\em Introduction.---} Many applications in quantum technology
require measurements
that are fast,
 accurate,
 and non demolition
 (in the sense that they do not induce transitions between states of the system one is trying to measure). In some cases,
 an  additional requirement arises when the system being measured is composed of multiple strongly-interacting sub-units.
If we wish to access information beyond the composite eigenstructure of that system, we often require a measurement device that itself is strongly coupled to the composite structure.

This issue has been analyzed recently in the context of ultrastrong coupling \cite{Niemczyk2010,Ashhab2010,Forn-Diaz2010,Beaudoin2011,PhysRevLett.105.196402,PhysRevB.79.201303,Scalari1323,PhysRevLett.117.043601,Feynmann,PhysRevA.92.063830,PhysRevA.88.063829,Diaz2016,Chen2016,Yoshihara2016} between light and matter in cavity QED \cite{Kavokin,Dutra}.
In such systems, the virtual photon occupation of the dressed ground state can be investigated by a non-adiabatic modulation of the interaction between light and matter  \cite{PhysRevB.72.115303,Sub-cycle,PhysRevLett.106.013601} or by inducing transitions outside the system's interacting Hilbert space \cite{Cirio2016,PhysRevLett.110.243601,PhysRevA.80.053810,RevModPhys.84.1}.\\

A natural step further is to look for methods which use minimal amount of resources \cite{Felicetti}, or are minimally invasive \cite{Lolli}, to study the dressed structure of the system. For example, in \cite{Lolli}, an ancillary qubit is used to investigate the ground state without disruptively disturbing it. Here, we consider an alternative method by using a hybrid matter-cavity-mechanical device \cite{Aspelmeyer2014,bowen2015quantum, Florian,Schliesser,PhysRevA.92.033806,PhysRevA.93.022510,Sun2009,PhysRevA.92.023846,
RevModPhys.85.623,BMT,PhysRevA.49.433,PhysRevLett.112.013601,PhysRevLett.114.173602,Rouxinol2016,Restrepo2016} where a mechanical mode, acting as the probe, couples via radiation pressure to a cavity-QED system (in which resonant matter ultrastrongly  interacts with the confined light). While it is clear that photons dressing the ground state of the strongly-coupled cavity-QED system can displace the mechanical mode through a ``virtual radiation pressure''  \cite{PhysRevA.38.1621,Dutra} effect (akin to variations of the Casimir force experiment \cite{PhysRevLett.78.5,MilonniBook,PhysRevLett.103.147003,PhysRevA.82.052509,WilsonNature}), typically, such a force is extremely weak.  Even if an exceptionally large opto-mechanical coupling was engineered to improve the measurement, virtual excitations might remain unobservable as the entire probe and light-matter system would relax to a combined collective ground state. Here we show that, even with a relatively weak opto-mechanical probe interaction strength, a modulation of the cavity-mechanical probe (i.e., opto-mechanical) interaction itself at the probe frequency can amplify the transduction of these virtual excitations into an observable displacement of the mechanical probe.

We begin with a description of the composite system, part by part, and intuitively derive the requirements for the detection of virtual radiation pressure effects with such a mechanical probe at zero temperature. We then give an analytical quantitative analysis, which includes thermal noise affecting both light-matter and mechanical systems.  As a result, we estimate the strength of the opto-mechanical coupling, and the bounds on the thermal noise, needed to resolve the effect within the standard quantum limit. Finally, we outline several explicit physical systems in which our proposal could be realized. \\

\emph{ Ultra-strong coupling of light and matter.---} The interaction between (a mode of) light confined in a cavity and a matter degree of freedom (modelled as a two-level system) is described by the quantum Rabi model \cite{PhysRev.49.324,Rabi1937} ($\hbar =1$),
\begin{equation}
\label{eq:H1}
H_\text{R}= \omega a^\dagger a +  \frac{\omega}{2}\sigma_z+ \Omega (\sigma_++\sigma_-) (a+a^\dagger)\;\;,
\end{equation}
where the fundamental mode of the cavity, with frequency $\omega$, is described by the annihilation operator $a$, the two-level system (assumed resonant with the cavity) is described by the Pauli operator $\sigma_z$. In this model, the light-matter interaction is fully  characterized by the normalized coupling $\eta\equiv{\Omega}/{\omega}$. In the weak-coupling regime ($\eta\ll 1$), terms which do not conserve the total free  excitation number can be neglected, leading to the Jaynes-Cummings interaction \cite{Jaynes1963}.  Therefore, the ground state $\ket{G}$ of the system does not contain photons, i.e., $\bar{n}_\text{GS}=\bra{G}a^\dagger a\ket{G}=0$. However, in the ultrastrong-coupling regime ($\eta>0.1$) hybridization effects play an important role and these qualitatively change the nature of the ground state (GS) which becomes dressed by virtual photons;  e.g., second-order perturbation theory in $\eta$ implies
\begin{equation}
\label{eq:ngs}
\bar{n}_{\text{GS}}=\bra{G}a^\dagger a\ket{G}\approx \frac{\eta^2}{4}\;\;.
\end{equation}
 Importantly, when weakly coupled to a low-temperature environment, the  system relaxes to the hybridized ground state $\ket{G}$, out of which photons cannot escape \cite{Beaudoin2011, breuer2002theory, PhysRevLett.109.193602}. As mentioned in the introduction, to observe such virtual excitations, we now introduce the concept of a mechanical probe, and show how active modulation of the probe's interaction with the above system allows for an amplified measurement of the ground state occupation.

\emph{Opto-mechanical interaction.--} The opto-mechanical interaction of a mechanical probe with the light-matter system described above can be most easily understood through the picture of a Fabry-Perot cavity with a mechanically-compliant mirror coupled to a spring with frequency $\omega_m$. This frequency is usually much smaller than the cavity frequency $\omega$. The interaction between photons inside the cavity and the mirror displacement is essentially radiation pressure, i.e., momentum kicks on the mechanical spring due to the bouncing of photons off the mirror. It  can be described, to lowest order in the displacement of the mirror, as
\begin{equation}
\label{eq:Ham}
H=H_R+ \omega_m b^\dagger b+g_0 a^\dagger a (b+b^\dagger)\;\;,
\end{equation}
where  $b$  is the annihilation operator of the mechanical mode, and  $g_0$ is the vacuum opto-mechanical coupling strength.  Note that, when matter is within the cavity, a third-order interaction term can arise because of modulation of the light-matter coupling strength $\Omega$ as the cavity length varies in time \cite{Sun2009}. Here we neglect that interaction, as it can be made negligible (while still maintaining a strong light-matter dipole coupling) by moving the position of the matter inside the cavity slightly away from the maximum of the electric field. Thus, here we focus on the standard opto-mechanical interaction term, for which the coupling amplitude $g_0$ corresponds to the frequency shift of the cavity when the mechanical displacement is equal to its zero-point motion $x_{zp}$  \cite{bowen2015quantum}.  Because of this interaction, in the absence of matter, an average of $n$ photons in the cavity exerts a radiation-pressure force $P_{n}=ng_0/x_{\text{zp}}$ on the mirror, inducing a displacement
\begin{equation}
\label{eq:average_intr}
|\langle x\rangle_{n}|=2n\eta_m x_{zp}\;\;,
\end{equation}
as a function of the normalized opto-mechanical coupling $\eta_m\equiv g_0/\omega_m$. Let us now provide some intuition on how the situation changes when an atom interacts with the cavity field. At sufficiently low temperatures, the cavity-QED composite system is in its ground state which still exerts a (virtual) radiation pressure on the mirror, readily found by setting $n=\bar{n}_\text{GS}$, giving
\begin{equation}
\label{eq:intuitive}
|\langle x\rangle_\text{GS}|=\frac{\eta^2}{2}\eta_m x_{zp}\;\;.
\end{equation}
To resolve the effect within the standard quantum limit, we need to impose $|\langle x\rangle _\text{GS}|>x_{zp}$, which leads to
\begin{equation}
\label{eq:requirement}
\eta_m>\frac{2}{\eta^2}\;\;.
\end{equation}
While  it is now possible for many different cavity-QED systems to reach the ultrastrong coupling regime $\eta\sim 0.1$, most realizations of opto-mechanical systems tend to be in the weak coupling regime $\eta_m\ll 1$, limiting the practicality of Eq.~(\ref{eq:requirement}) (although proposals to achieve stronger couplings do exist \cite{PhysRevLett.114.143602,PhysRevA.93.022510,1367-2630-16-5-055008,PhysRevLett.112.203603,PhysRevLett.116.163602,PhysRevLett.114.093602,Xuereb1,Xuereb3}).

 However, we can overcome this limitation by modulating the opto-mechanical coupling $g_0\mapsto g_0(t)$, akin to recent proposals to enhance effective Kerr nonlinearities with a modulated opto-mechanical coupling \cite{LiaoJacobs2014}, to enhance the readout of qubits  with a modulated  longitudinal coupling \cite{Royer2016}, or by modulating other parameters of the atom-cavity system \cite{PhysRevA.93.063845,PhysRevA.91.063814}.  Intuitively, this modulation effectively turns radiation pressure into a built-in (photon-number-dependent) resonant driving force. With this interpretation in mind, by considering a modulation at the mechanical frequency
 \begin{equation}
 \label{eq:modulation}
 g_0\mapsto g_0\cos{\omega_m t}\;\;,
 \end{equation}
  we immediately find \footnote{Intuitively, the  mechanical displacement induced by the  (Fourier transformed) radiation pressure force $P_{n}(\omega)$ can be written as $\Delta x=\int_{-\infty}^\infty\!d\omega~ e^{i\omega t}\chi(\omega)P_{n}(\omega)$ in terms of the mechanical susceptibility $\chi(\omega)=\omega_m/(\omega_m^2-\omega^2-i\omega\Gamma_m)$. For constant opto-mechanical couplings we have $P_{n}\propto\delta(0)$. By modulating the coupling $g_0$ at the mechanical frequency, the radiation pressure effectively becomes a driving force with $P_{n}\propto(\delta(\omega-\omega_m)+\delta(\omega+\omega_m))/2$. By inserting these expression in the one for $\Delta x$, we then find (in a rotating frame) the  enhancement factor reported in the main text.} that the mechanical displacement is enhanced by the factor $|\chi(\omega_m)|/|\chi(0)|=\omega_m/\Gamma_m$  in terms of the frequency-dependent mechanical susceptibility $\chi(\omega)$ and the mechanical decay rate $\Gamma_m$.  This effectively corresponds to the substitution $\eta_m\mapsto \bar{\eta}_m$, (with $\bar{\eta}_m=g_0/\Gamma_m$) in Eq.~(\ref{eq:requirement}), obtaining the much more realistic requirement
\begin{equation}
\label{eq:condition}
\bar{\eta}_m>\frac{2}{\eta^2}\;\;.
\end{equation}
 This suggests the amplified observation of ground-state excitations is feasible, and constitutes our first main result.  While this result holds for zero temperature, at small but finite temperatures, correlations between the system and the mechanical probe arise, which can complicate the problem of distinguishing the small thermal occupation of the light-matter system from virtual excitations.

 To understand in detail the competition between ground state occupation and unwanted environmental influence, we perform a detailed  analysis, based on an analytical low-energy effective model. This allows us to estimate temperature-dependent bounds for the observation of the virtual excitations.
In addition, we will show that the protocol presented here does not amplify the intrinsic mechanical thermal noise, which we expect to be the most relevant in realistic implementations (wherein the mechanical probe frequency is much smaller than the strongly-coupled light-matter parameters).

\emph {Effective model.---} With the modulation of the opto-mechanical coupling  described in Eq.~(\ref{eq:modulation}), and  in a frame rotating at the mechanical frequency $\omega_m$ the Hamiltonian in  Eq.~(\ref{eq:Ham}) becomes
\begin{equation}
\label{eq:Hmodulated}
H= H_R+\frac{g_0}{2}a^\dagger a(b+b^\dagger)\;\;,
\end{equation}
where we performed a rotating wave approximation (see Appendix for the non-resonant driving case).

A Born-Markov perturbative master-equation treatment of the interaction with the environment for the system in Eq.~(\ref{eq:Hmodulated}) %\ref{eq:Ham}
can be written as $\dot{\rho}=-i[H,\rho]+\mathcal{L}_R(\rho)+\mathcal{L}_m(\rho)$ \cite{breuer2002theory,Carmichael2002}, where
 the term $\mathcal{L}_m= \Gamma_m\left(\bar{n}_m\mathcal{D}[b^\dagger](\rho)+(1+\bar{n}_m)\mathcal{D}[b](\rho)\right)$ is the Liouvillian, accounting for the bath of the mechanical degree of freedom, as a function of its thermal occupation number $\bar{n}_m$ and where $\mathcal{D}[O](\rho)=\frac{1}{2}(2O\rho O^\dagger-\rho O^\dagger O-O^\dagger O \rho)$. The Liouvillian $\mathcal{L}_R$ depends on the environments coupled to the photonic and matter systems and, importantly, in the ultrastrong coupling regime, causes transitions between dressed states which diagonalize the light-matter Hamiltonian~\cite{breuer2002theory,Beaudoin2011}.
We now assume a regime where the population of the light-matter system is restricted to its lowest (dressed) energy states, i.e. the ground $\ket{G}$ and first two excited states $\ket{\pm}$. Under this approximation, we can project $H$ to this low-energy subspace. Not surprisingly, in this limit, the model can be given a bosonic representation under the replacement $\ketbra{G}{\pm}\mapsto a_\pm$, where the bosonic annihilation operators $a_\pm$ now carry information about the low-energy structure of the light-matter Hilbert space. In this way, it is possible to provide an analytical treatment of the model, including a self-consistent quantification of the low-temperature effects. Under these assumptions, to second order in $\eta$, the Hamiltonian reads
\begin{equation}
\label{eq:222}
H=\omega_+a^\dagger_+a_++\omega_-a^\dagger_-a_-+\frac{g_0}{2} \hat{\alpha}(b+b^\dagger)\;\;,
\end{equation}
where $\omega_\pm=\omega (1\pm\eta)$ , $\hat{\alpha}=(\alpha_+ a^\dagger_+a_++\alpha_-a^\dagger_-a_-+\xi)$, with $\alpha_\pm= \frac{1}{2}\mp{\eta}/{4}$, $\xi={\eta^2}/{4}$, and where we neglected terms rotating at frequencies $2\omega$ and $2 \eta \omega$ in the opto-mechanical interaction term.
In addition, this result enlarges the domain of our analysis to physical systems with {\em a priori} bosonized matter degrees of freedom (as is typical for many-particle systems like quantum wells). Indeed, by replacing $\sigma_-$ in Eq.~(\ref{eq:H1}) with the annihilation operator of a harmonic mode, the system can be diagonalized by a standard Bogoliubov transformation and takes the same form  as Eq.~(\ref{eq:222}) with re-defined parameters (\cite{Supplemental}, section III).
Thus, all the results given below and written as a function of these coefficients are valid for both the spin and bosonic cases.

In the linearized approximation we are considering, a completely equivalent master equation for the coupled system can naturally be written (\cite{Supplemental}, section II) in terms of three independent baths as
\begin{equation}
\label{eq:masterEq}
\begin{array}{lll}
\dot{\rho}=-i[H,\rho]+\mathcal{L}_+(\rho)+\mathcal{L}_-(\rho)+\mathcal{L}_m(\rho)\;\;,
\end{array}
\end{equation}
where $\mathcal{L}_\pm(\rho)= \kappa_\pm\left(\bar{n}_\pm\mathcal{D}[a_\pm^\dagger](\rho)+(1+\bar{n}_\pm)\mathcal{D}[a_\pm](\rho)\right)$,  and $\kappa_\pm$ are linear combinations of the decay rates of the light-matter subsystems calculated at the  frequencies $\omega_\pm$.
For simplicity, in the following we will assume that the occupation numbers are equal $\bar{n}_\pm=\bar{n}$ (see \cite{Supplemental} for details and a more general analysis).\\
\begin{figure}[t!]
\centering
\includegraphics[width=\linewidth]{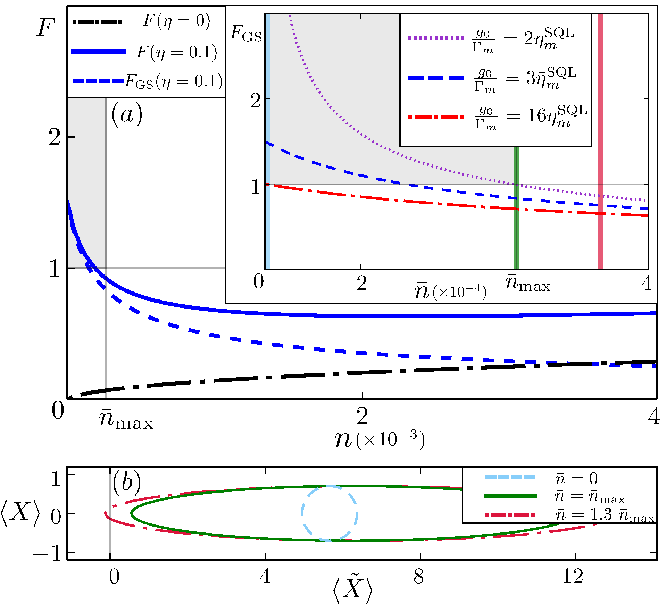}
\caption{\label{Figgg}(Color online) (a):
 Total displacement visibility $F$ in the presence (full blue curve, $\eta=0.1$) and absence (full black curve, $\eta=0$) of matter in the cavity as a function of the number of  thermal light-matter excitations $\bar{n}$ (for an optomechanical coupling $g_0/{\Gamma_m}=3\bar{\eta}_0^{\text{SQL}}$, for $\bar{\eta}_0^{\text{SQL}}=2/{\eta^2}$). For high values of $\bar{n}$ the two curves asymptotically converge to a parallel behaviour. In the absence of matter,  when $\bar{n}\rightarrow 0$ a zero photon population implies no  displacement (black curve). However, in the presence of matter, virtual photons can displace the mechanical oscillator even for $\bar{n}\rightarrow 0$ (blue curve).
The relative displacement contribution purely due to virtual radiation pressure effects $F_\text{GS}$ is represented by the blue dashed curve showing that, for $\bar{n}\rightarrow 0$, the displacement is mainly due to the dressed structure of the ground state. The grey vertical line represents the theoretical upper bound $\bar{n}_\text{max}$. Below this critical value, it is possible to tune $g_0/{\Gamma_m}$ to resolve the ground state signal. This is shown in the Inset which magnifies the main plot around  $\bar{n}=\bar{n}_\text{max}$. The blue curve corresponds to the same color coded ones in the main figure (a). The dotted purple and dot-dashed red curves are plotted for different values of $g_0/{\Gamma_m}$ ($16\bar{\eta}_0^{\text{SQL}}$ and $2\bar{\eta}_0^{\text{SQL}}$, respectively). For $\bar{n}<\bar{n}_\text{max}$, it is always possible to find opto-mechanical couplings which, in principle, allow one to resolve the ground state signal (i.e., $F_\text{GS}>1$). At values of $\bar{n}$ highlighted in the inset, in (b) we plot (following the color code) a phasor diagram for the state of the system, i.e. its quadrature displacement (for $X=(b+b^\dagger)/{\sqrt{2}}$ and $\tilde{X}=i(b^\dagger-b)/{\sqrt{2}}$) and standard deviation. This for  $g_0/{\Gamma_m}=16\bar{\eta}_0^{\text{SQL}}$, i.e., the one corresponding to the violet curve in the inset. While for $\bar{n}=0$ the displacement above zero can be clearly resolved, this visibility decreases as we approach and overtake the value $\bar{n}_\text{max}$.}
\end{figure}

\emph{Enhanced readout.---} From Eq.~(\ref{eq:222}),  note that the force acting on the mechanical mode $P= g_0 \hat{\alpha}/2x_\text{zp}$ has two contributions: the usual radiation pressure (dependent on the number of normal excitations in the light-matter system) and virtual radiation pressure (proportional to $\xi$, accounting for ground state effects). Following Eq.~(\ref{eq:masterEq}), the Heisenberg equation of motion  for the dimensionless quadrature of the mechanical mode $|\langle \tilde{X}\rangle|={|\langle i( b^\dagger-b)\rangle|}/{\sqrt{2}}$ in the steady state leads to
\begin{equation}
\label{eq:average}
|\langle \tilde{X}\rangle|=\sqrt{2}\bar{\eta}_m(\alpha_+ \bar{n}_++\alpha_-\bar{n}_- +\xi),
\end{equation}
which is the expected result from our intuitive analysis in the introduction: the modulation of the coupling induces a displacement of the mechanical probe with an amplified amplitude proportional to $\bar{\eta}_m={g_0}/{\Gamma_m}$. As implicitly done throughout the article, we omitted zero-point energy contributions  \footnote{A constant displacement due to zero-point energy contributions has been subtracted from the quadrature average. Importantly, the variances do not depend on such contributions \cite{Supplemental} (section III). As a consequence, they play no role in the following analysis.}. As shown by this expression,  the total displacement has two physically-different contributions, i.e., $|\langle \tilde{X}\rangle_\text{GS}|=({\xi}/{\alpha})|\langle \tilde{X}\rangle|$ (accounting for virtual radiation-pressure effects) and $|\langle \tilde{X}\rangle_{\bar{n}}|=({1-\xi/\alpha})|\langle \tilde{X}\rangle|$ (accounting for finite temperature effects), where $\alpha=\langle\hat{\alpha}\rangle=\alpha_+ \bar{n}_++\alpha_-\bar{n}_- +\xi$.

\emph{Signal-to-noise ratio.---} To analyse the interplay between the two different contributions to the displacement and to what degree they can be resolved, both from one another and from the mechanical systems own vacuum fluctuations (the standard quantum limit), we use the ratio $F\equiv{|\langle \tilde{X} \rangle|}/{\delta\!\tilde{X}}$, where  $(\delta\!\tilde{X})^2=\langle \tilde{X}^2\rangle-\langle \tilde{X}\rangle^2$, a general analytical expression of which is shown in \cite{Supplemental} (section III.B).  At finite temperatures, the mechanical probe and the light-matter system become correlated, leading to a non-trivial expression for this variance.  Using  Eq.~(\ref{eq:average}), we can define the analogous ratio for the ground state signal contribution alone as
\begin{equation}
F_{\text{GS}}\equiv\frac{|\langle \tilde{X} \rangle_\text{GS}|}{\delta\!\tilde{X}}\;\;,
\end{equation}
which quantifies our ability to resolve virtual radiation pressure effects.
We plot  \cite{Johansson20121760,Johansson20131234} these quantities as a function of the thermal occupation of the light-matter system in Fig.~\ref{Figgg}. There, we show how the behaviour of $F$ is qualitatively different in the presence (blue curves) or absence (black curve) of matter in the cavity. For higher occupation numbers the curves asymptotically converge to a $\eta$-dependent constant, as expected when thermal effects dominate (\cite{Supplemental}, section III.B). However, close to the ground state, a null value of $F$ in the absence of matter gives $F\rightarrow F_{\text{GS}}\neq 0$ (dashed blue curve) when matter is present in the cavity in the ultrastrong-coupling regime.

For a more quantitative analysis, we now consider two minimal conditions to observe the influence of virtual radiation pressure on the mechanical displacement, i.e. the conditions
\begin{equation}
\label{eq:requirements}
\begin{array}{lllllll}
|\langle \tilde{X}\rangle_{\text{GS}}|&>& |\langle \tilde{X}\rangle_{\bar{n}}|,&\qquad\qquad &F_{\text{GS}}&>&1\;\;.
\end{array}
\end{equation}
The first condition requires the observed total displacement to be mainly due to ground state effects.
The second condition requires the signal to be resolved with respect to the standard-quantum-limit noise \cite{Braginsky,RevModPhys.82.1155} (see threshold in Fig. \ref{Figgg}).

From the analysis following Eq.~(\ref{eq:average}), the first condition translates to an upper bound $\bar{n}_\text{GS}$ on the allowed thermal occupation of the light-matter system for the ground-state effects to dominate. Complementarily, the second condition implies the ability to resolve the ground-state contribution to the signal in Eq.~(\ref{eq:average}) with respect to its total uncertainty $\delta\!\tilde{X}$. It translates into both a lower bound  ${\bar{\eta}^{\text{SQL}}}_m$ on the normalized opto-mechanical coupling and another upper bound $\bar{n}^{\text{SQL}}$ on the thermal light-matter occupation. By solving the Heisenberg equation of motion using Eq.~(\ref{eq:masterEq}), we  find \cite{Supplemental} the following explicit conditions
\begin{equation}
\label{eq:allCond}
\bar{n}<\bar{n}_\text{max},\qquad\qquad  \bar{\eta}_m>\bar{\eta}^\text{SQL}_m\;\;.
\end{equation}
This is the second main result of our work, generalizing \equ{eq:condition} to finite temperatures. Here,  $2\bar{\eta}^\text{SQL}_m=[({{1+2n_b})/{(\xi^2-R)}}]^{1/2}$ (with $R=\bar{n}(1+\bar{n})({\alpha^2_+}/{{\beta_+}}+{\alpha^2_-}/{{\beta_-}})$, $\beta_\pm=1+2{\kappa_\pm}/{\Gamma_m}$) and $\bar{n}_\text{max}=\text{min}\left(\bar{n}^\text{GS},\bar{n}^{\text{SQL}}\right)$ (with  $4\bar{n}^{\text{GS}}={\eta^2}$, $8\bar{n}^{\text{SQL}}={\beta}\eta^4$ at lowest significant order in $\eta$ where the expression for $n_\text{max}$ does not depend on the bosonic or spin nature of the model).

Consistent with our initial intuitive reasoning, when $n_b, R\rightarrow 0$, the second expression in Eq.~(\ref{eq:allCond}) is equivalent to the zero-temperature result given in Eq.~(\ref{eq:condition}). Moreover, we note that mechanical thermal occupation is not amplified by this protocol,
and its  influence can be understood as a weak renormalization of the  opto-mechanical coupling $g_0\mapsto{g_0}/{\sqrt{1+n_b}}$.  In summary, one can observe the amplified ground-state occupation when the temperature is low enough such that ground-state effects both dominate the displacement ($\bar{n}<\bar{n}_\text{GS}$) \emph{and} can be resolved from thermal and vacuum fluctuations (which requires $\bar{n}<\bar{n}^\text{SQL}$, and sufficiently large opto-mechanical coupling $\bar{\eta}_m>\bar{\eta}_m^\text{SQL}$).

The dependence of the signal-to-noise ratio on the thermal noise is shown in Fig. \ref{Figgg} (b), with the quadrature displacements and their variance at different temperatures and for a fixed value of the opto-mechanical coupling. For $\bar{n}>\bar{n}_{\text{max}}$, the average displacement increases but  it is only due to thermal noise. However, for $\bar{n}<\bar{n}_{\text{max}}$, the displacement is mainly due to virtual radiation-pressure effects and it can in principle be resolved.

{\em Experimental feasibility.---} In the optical regime, a membrane-cavity system, or a photonic crystal cavity, could realize our proposal. For example, in a typical example of a membrane-cavity system, the membrane  (acting as the mechanical probe)  lies within the fixed mirrors of an optical cavity \cite{Harris,PhysRevA.92.013822,1367-2630-10-9-095008,PhysRevLett.109.063601,PhysRevA.84.023812,PhysRevA.77.033819,Xuereb2}. Effectively, the membrane splits the cavity in two, whose frequencies depend on the position of the oscillator. By tuning the position of the membrane with respect to the modes of the cavities, it is possible to obtain either a linear or quadratic opto-mechanical coupling. Consequently, an effective modulation of the opto-mechanical coupling can be obtained by modulating the initial position $q_0$ in time \cite{PhysRevA.92.013822}. The advantage of this implementation is that works in the optical regime where $\bar{n}\sim 0$.

There are also many opto-mechanical devices which operate at  microwave frequencies. The advantage of these schemes lies in the possibility to more easily achieve stronger electro-mechanical couplings $g_0$ and well as ultrastrong light-matter interaction ($\eta>0.1$, \cite{Yoshihara2016}). The most well-known example is that of a microwave cavity capacitively coupled to a micro-mechanical membrane \cite{LiaoJacobs2014,Teufel2011}. The experimental parameters realized in these systems are very promising, with the thermal occupation of the cavity being just $\bar{n}\sim 10^{-10}$ and renormalized opto-mechanical coupling  $\bar{\eta}_m\sim 5$  which, together with the possibility of light-matter coupling reaching $\eta>0.1$ in circuit-QED devices \cite{Yoshihara2016}, would allow one to fulfill Eq.~(\ref{eq:condition}).

 Another possibility would be to implement the proposal given in  \cite{PhysRevA.93.022510}  where  a SQUID with a mechanically-compliant arm is coupled to a coplanar microwave cavity. Alternatively, also in the microwave regime, one could replace the mechanical probe with an equivalent microwave cavity probe  by having two microwave resonators interact via a SQUID-mediated opto-mechanical-like interaction \cite{Johansson2014,Kim2015}. This approach would allow a stronger modulated opto-mechanical-like coupling.

In these examples, if a time-dependent coupling is difficult to realize, using a parametric driving of the mechanical mode  could effectively lead to a time-dependent opto-mechanical coupling \footnote{A parametric modulation of the mechanical frequency itself \cite{Zhukov2016, Lemonde2016} can lead to an effective model with a time-dependent coupling. This will be fully analyzed in a future work \cite{InProgress}.}.

{\em Conclusions.---} In this article we presented a method to probe the structure of the dressed ground state by introducing  an ``opto-mechanical" coupling between the cavity mode and a measurement device (which may either be a real mechanical device \cite{PhysRevLett.107.223001,PhysRevA.82.021803}, or an artificial physical simulation, e.g., a second microwave cavity engineered to have an opto-mechanical coupling \cite{Kim2015,PhysRevA.93.022510,Johansson2014}).
  Normally, the sensitivity of  such a device to the presence of virtual photons is limited by the strength of the light-matter interaction (quantified by $\eta$) and the weak nature of the opto-mechanical coupling. We showed that a time-dependent modulation of the opto-mechanical coupling leads to an effective amplification of the measurement strength, allowing one to peer into the dressed ground state.  We expect that this technique could also be applied to other measurement problems based on the same opto-mechanical interaction.

{\em Acknowledgments.---} We acknowledge helpful feedback and comments from Anton F. Kockum, Simone De Liberato,  Adam Miranowicz, Roberto Stassi, and Jacob Taylor. FN acknowledges support from the RIKEN iTHES Project, the MURI Center for Dynamic Magneto-Optics via the AFOSR award number FA9550-14-1-0040, the IMPACT program of JST, CREST, and a Grant-in-Aid for Scientific Research (A). FN, MC, and NL acknowledge support from the Sir John Templeton Foundation.

\bibliography{Ref}

\include{Supplemental_content.tex}

\end{document}

\newpage
%\onecolumngrid
\newpage

%\section*{\LARGE{Supplemental Material}}
%\onecolumngrid
%\maketitle
\title{Supplemental Material for\\ Amplified opto-mechanical transduction of virtual radiation pressure}
\author{Mauro Cirio}
\affiliation{CEMS, RIKEN, Wako-shi, Saitama 351-0198, Japan}

\author{Kamanasish Debnath}
\affiliation{CEMS, RIKEN, Wako-shi, Saitama 351-0198, Japan}
\affiliation{Amity Institute of Applied Sciences, Amity University, Noida - 201303 (U.P.), India}
\affiliation{Laboratory of Theoretical Physics of Nanosystems, Ecole Polytechnique F\'ed\'erale de Lausanne (EPFL), CH-1015 Lausanne, Switzerland}

\author{Neill Lambert}
\affiliation{CEMS, RIKEN, Wako-shi, Saitama 351-0198, Japan}

\author{Franco Nori}
\affiliation{CEMS, RIKEN, Wako-shi, Saitama 351-0198, Japan}
\affiliation{Department of Physics, University of Michigan, Ann Arbor, Michigan 48109-1040, USA}
\begin{abstract}
In this supplemental material we present the details of the hybrid opto-mechanical structure considered in the main text. The system consists of an electromagnetic mode ultrastrongly coupled to a matter degree of freedom (via dipole interaction) and to a mechanical oscillator (via radiation pressure). In this way, the mechanical mode can be used as a probe of the dressed structure of the light-matter system, i.e., as a transducer of virtual radiation pressure. To amplify the signal, we consider a modulation of the opto-mechanical interaction at the mechanical frequency. We model the matter as either a spin (in the low-energy limit) or a bosonic mode and find a unified effective master equation (with different parameters) which describe them. We use these results to calculate bounds on the minimal amount of resources necessary to resolve virtual radiation-pressure effects when probing the mechanical quadratures.
\end{abstract}
\maketitle
\onecolumngrid
\section{Low-energy bosonized light-matter-mechanical model}
\label{AppA}
In this section, we derive the effective Hamiltonian to describe the low-energy physics of opto-mechanical systems where the cavity mode interacts ultrastrongly with a two-level atom (spin). In turn, such a model can be used to exactly describe the case where the matter degree of freedom can be directly modelled as bosonic. For symmetry, throughout this supplemental material the cavity mode has been relabelled as $a\mapsto a_1$.
\subsection{Spin case}
We consider the standard physical situation in which a spin (described by the operator $\sigma_\pm$) resonantly interacts with an electromagnetic mode confined in a cavity (described by the operator $a$) whose frequency $\omega$ is modulated by the position $x=x_{\text{zp}}(b+b^\dagger)$ of the mechanical mode $b$ of frequency $\omega_m$. The Hamiltonian can be written as \cite{Aspelmeyer2014,bowen2015quantum}
\begin{equation}
H = \omega(x) a_1^\dagger a_1 + \frac{\omega}{2} \sigma_z + \Omega (a_1^\dagger + a_1)(\sigma_++\sigma_-) + \omega_m b^\dagger b\;\;.
\end{equation}

As mentioned in the main text, third-order interaction terms can arise from the modulation of the field strength at the atom position as a consequence of the mechanical motion. However, such contributions can be made negligible by tuning the position of the atom inside the cavity while still being close to the point of maximum intensity of the electric field  \cite{Sun2009}. Now, by expanding to first order in $x$, we obtain
\begin{equation}
\label{eq:1}
H = H_R+\omega_m b^\dagger b + g_0 a_1^\dagger a_1 (b+b^\dagger)\;\;,
\end{equation}
with the vacuum opto-mechanical coupling
\begin{equation}
g_0 = x_{\text{zp}}\frac{\partial \omega}{\partial x}_{|_{x=0}}\;\;,
\end{equation}
and where the light matter system is described by the quantum Rabi Hamiltonian
\begin{equation}
H_\text{R}= \omega a_1^\dagger a_1 +  \frac{\omega}{2}\sigma_z+ \Omega (\sigma_++\sigma_-) (a_1+a_1^\dagger)\;\;.
\end{equation}
Here, we omitted zero-point energy contributions. We further considered the effect of this omission in section III.\\
The ultrastrong coupling regime for the light-matter system (occurring when the normalized Rabi coupling
\begin{equation}
\eta=\frac{\Omega}{\omega}\geq 0.1\;\;,
\end{equation}
 implies a perturbative characterization of the environment \cite{breuer2002theory,Beaudoin2011} which induces transitions between Rabi dressed eigenstates. For this reason, we define a low-energy effective model by
 \begin{equation}
 H^{\text{eff}}= P H P\;\;,
 \end{equation}
 where $P$ is the projector into a low energy sector of the full Hilbert space
 \begin{equation}
 P=\ketbra{G}{G}+\ketbra{-}{-}+\ketbra{+}{+}\;\;,
 \end{equation}
  where $\ket{G}, \ket{\pm}$ are the three energy eigenstates of the Rabi Hamiltonian $H_R$ with lowest energy.  By using quasi-degenerate perturbation theory, at second order in  $\eta$, these states are found to be
\begin{equation}
\label{eq:fstates}
\begin{array}{lll}
\ket{G}&=&(1-\frac{\eta^2}{8})\ket{0,g}+\frac{\eta}{2\sqrt{2}}(\ket{2,-}-\ket{2,+})+\frac{\eta^2}{4}(\ket{2,-}+\ket{2,+})\\
\ket{1,\pm}&=&(1-\frac{\eta^2}{8}-\frac{\eta^2}{32})\ket{1,\pm}\mp\frac{\eta}{4}(1\pm\frac{\eta}{2})\ket{1,\mp}+(\frac{\eta}{2\sqrt{2}}\pm\frac{\eta^2}{8\sqrt{2}})(\ket{3,-}-\ket{3,+})+\frac{\sqrt{3}\eta^2}{4\sqrt{2}}(\ket{3,-}+\ket{3,+})\;\;,
\end{array}
\end{equation}
where
\begin{equation}
{\renewcommand{\arraystretch}{1.2}
\begin{array}{lll}
\ket{0}&=&\ket{0,g}\\
\ket{n,\pm}&=&\frac{\ket{n,g}\pm\ket{n-1,e}}{\sqrt{2}}\;\;,
\end{array}}
\end{equation}
for  $n\geq 1$ and where $a_1\ket{0,g}=\sigma^-\ket{0,g}=0$, and $a_1^\dagger a_1\ket{n,g}=n\ket{n,g}$, $a_1^\dagger a_1\ket{n,e}=n\ket{n,e}$ and $\sigma_z\ket{n,g}=-\ket{n,g}$, $\sigma_z\ket{n,e}=\ket{n,e}$. At the same order, the corresponding energies are
\begin{equation}
\frac{\tilde{E}_0}{\omega}=-\frac{1}{2}-\frac{\eta^2}{2}+O(\eta^3)\;\;,\qquad\qquad\qquad
\frac{\tilde{\omega}_\pm}{\omega}=\frac{1}{2}\pm\eta-\frac{\eta^2}{2}+O(\eta^3)\;\;.
\end{equation}

Moreover, we have the following identities
\begin{equation}
{\renewcommand{\arraystretch}{1.75}\begin{array}{lllll}
\xi&\equiv&\bra{G}a_1^\dagger a_1\ket{G}&=&\frac{\eta^2}{4}\\
\tilde{\alpha}_\pm&\equiv&\bra{\pm}a_1^\dagger a_1\ket{\pm}&=&\frac{1}{2}\mp\frac{\eta}{4}+\frac{\eta^2}{4}\\
&&\bra{\pm}a_1^\dagger a_1\ket{\mp}&=&\frac{1}{2}+\frac{3}{16}\eta^2\\
&&\bra{G}a_1^\dagger a_1\ket{\pm}&=&0\;\;.
\end{array}}
\end{equation}

 These results allow us to write
\begin{equation}
\begin{array}{lll}
 H^\text{eff}&=&H^\text{eff}_R+g_0(b+b^\dagger) (\tilde{\alpha}_- \ketbra{-}{-}+\tilde{\alpha}_+\ketbra{+}{+}+\xi\ketbra{G}{G})+ \omega_m b^\dagger b\;\;,
\end{array}
\end{equation}
where
\begin{equation}
H^\text{eff}_R=\tilde{E}_0\ketbra{G}{G}+\tilde{\omega}_-\ketbra{-}{-}+\tilde{\omega}_+\ketbra{+}{+}\;\;,
\end{equation}
and where we omitted terms proportional to the operators $\ketbra{\pm}{\mp}$ in a rotating-wave approximation. The omission of these terms requires careful analysis. In fact, in the interaction picture described by the diagonalized Rabi Hamiltonian, these operators rotate at frequencies $\pm2\eta\omega$. However, the error produced by this approximation should not wash out the effect of the term  $g_0\xi (b+b^\dagger)\ketbra{G}{G}$, which is the one giving rise to the physics we are exploring. For example, by using second order Van-Vleck perturbation theory in Floquet space \cite{Leskes,Ernst,Alessio,Eckardt}, it is possible to show that, in a regime where ${g_0}/{\eta\omega}\ll 1$, the worst case errors are $O({g_0^2}/{\eta\omega})$ so that
\begin{equation}
\frac{g_0}{\omega}\ll\eta\xi\propto \eta^3\;\;,
\end{equation}
 is enough to justify this approximation. Interestingly, this procedure critically requires the ultrastrong coupling regime.

As routinely done in condensed matter physics, within the low-energy approximation considered here, we now map our model to a purely bosonic one. In turn, this will allow us to extend our analysis to physical systems where a bosonic approximation for matter degrees of freedom can be done \emph{a priori}, i.e., directly in the original Rabi Hamiltonian (see  next subsection). With this idea in mind, we first re-write the previous Hamiltonian as
\begin{equation}
 H^\text{eff}=\omega_-\ketbra{-}{-}+\omega_+\ketbra{+}{+}+g_0(b+b^\dagger) (\alpha_- \ketbra{-}{-}+\alpha_+\ketbra{+}{+}+\xi)+ \omega_m b^\dagger b\;\;,
\end{equation}
where
\begin{equation}
\begin{array}{lllll}
\omega_\pm&=&\tilde{\omega}_\pm-\tilde{E}_0&=&(1\pm\eta)\omega\\
\alpha_\pm&=&\tilde{\alpha}_\pm-\xi&=&(\frac{1}{2}\mp\frac{\eta}{4})\;\;,
\end{array}
\end{equation}
and where we used the fact that, within the low-energy sector of the Hilbert space the relation
\begin{equation}
\mathbb{I}=\ketbra{G}{G}+\ketbra{-}{-}+\ketbra{+}{+}\;\;,
\end{equation}
 holds as an effective identity. Now, the bosonization of the previous Hamiltonian can be carried on by imposing the substitution $\ketbra{G}{\pm}\mapsto a_\pm$ to get
\begin{equation}
 H^\text{eff}=\omega_-a^\dagger_-a_-+\omega_+a^\dagger_+a_++g_0(b+b^\dagger) (\alpha_- a^\dagger_- a_-+\alpha_+a^\dagger_+ a_++\xi)+ \omega_m b^\dagger b\;\;.
\end{equation}

By considering a modulation $g_0\mapsto g_0\cos{\omega_m t}$, by going to a frame rotating at the mechanical frequency \cite{LiaoJacobs2014,Royer2016,PhysRevA.93.063845,PhysRevA.91.063814}, and by subsequently performing a rotating wave approximation, we then exactly get Eq.~(10) as reported in the main text.
\subsection{Bosonic matter}
In a physical situation where the matter degree of freedom can be modelled as bosonic {\em a priori}, we can directly start our analysis by imposing the substitutions $\sigma_-\mapsto a_2$ in Eq.~(\ref{eq:1}), i.e., by replacing the spin with a harmonic oscillator with annihilation operator $a_2$. Explicitly
\begin{equation}
\label{eq:bb}
H^B = \omega a_1^\dagger a_1+\omega a_2^\dagger a_2+\Omega (a_1 +a_1^\dagger)(a_2+a_2^\dagger)+\omega_m b^\dagger b + g_0 a_1^\dagger a_1 (b+b^\dagger)\;\;.
\end{equation}
For $g_0=0$, this model is solvable and, by defining $a_\pm=({{m\omega_\pm}/{2}})^{1/2}[x_\pm+{i}/{(m\omega_\pm)}p_\pm]$, in terms of $x_\pm={x_1\pm x_2}/{\sqrt{2}}$, where $x_j=({2m\omega})^{-1/2}(a_j+a^\dagger_j)$ (j=1,2), we obtain, after some straightforward algebra

\begin{equation}
\label{eq:2}
\begin{array}{lll}
H^B&=&E_0+{\omega_+^B}a^\dagger_+a_++{\omega_-^B}a^\dagger_-a_-+\omega_m b^\dagger b+g_0(b+b^\dagger)\{\dfrac{1}{8\omega{\omega_+^B}}[(\omega^2-{\omega_+^B}^2)(a^\dagger_+a^\dagger_++a_+a_+)]
+\dfrac{1}{4\omega{\omega_+^B}}[(\omega^2+{\omega_+^B}^2)a^\dagger_+a_+]\\
&&+\dfrac{1}{8\omega{\omega_-^B}}[(\omega^2-{\omega_-^B}^2)(a^\dagger_-a^\dagger_-+a_-a_-)]
+\dfrac{1}{4\omega{\omega_-^B}}[(\omega^2+{\omega_-^B}^2)a^\dagger_-a_-]\}
+g_0(b+b^\dagger)[\dfrac{(\omega-{\omega_+^B})^2}{8\omega{\omega_+^B}}+\dfrac{(\omega-{\omega_-^B})^2}{8\omega{\omega_-^B}}]\\
&&+g_0(b+b^\dagger)\left\{\dfrac{1}{4\omega\sqrt{{\omega_+^B}{\omega_-^B}}}(\omega^2-{\omega_+^B}{\omega_-^B})(a_+^\dagger a_-^\dagger+a_+ a_-)\right.
\left.+\dfrac{1}{4\omega\sqrt{{\omega_+^B}{\omega_-^B}}}(\omega^2+{\omega_+^B}{\omega_-^B})(a_+^\dagger a_-+a_-^\dagger a_+)\right\}\;\;,
\end{array}
\end{equation}
where $E_0=({\omega^B_+ + \omega^B_-})/{2}-\omega$, $\omega^B_\pm=\omega({1\pm2\eta})^{1/2}$, and where, explicitly, the Bogoliubov relations reads
\begin{equation}
\label{eq:expr}
\begin{array}{lll}
a_j+a_j^\dagger=\sqrt{\dfrac{\omega}{2\omega^B_+}}(a_++a_+^\dagger)+(-1)^j\sqrt{\dfrac{\omega}{2\omega^B_-}}(a_-+a_-^\dagger)\;\;,
\end{array}
\end{equation}
for $j=1,2$.
We now notice that terms proportional to $a^\dagger_\pm a^\dagger_\pm$ and $a_\pm a_\pm$ rotate at frequencies $\pm 2\omega^B_\pm$ and can be neglected with a rotating-wave approximation. A similar analysis holds for the terms $a_+^\dagger a_-^\dagger$ and $a_+ a_-$, at lowest order in $\eta$. The terms proportional to $a^\dagger_+ a_-$ and $a_-^\dagger a_+$ rotate at a lower frequency (i.e., $O(\omega\eta)$) and their norm is suppressed by the factor $\eta^2$. For this reason, in order to neglect them, we need to carry the same perturbative considerations done in the spin case. In this approximation, we get our final result of this subsection
\begin{equation}
\label{eq:effective}
\begin{array}{lll}
H^B&=&\omega^B_+a^\dagger_+a_++\omega^B_-a^\dagger_-a_-+\omega_m b^\dagger b+g_0(b+b^\dagger)(\alpha^B_+a^\dagger_+a_++\alpha^B_-a^\dagger_-a_-+\xi^B)\;\;,
\end{array}
\end{equation}
with
\begin{equation}
{\renewcommand{\arraystretch}{2.5}
\begin{array}{lll}
\alpha^B_\pm&=&\dfrac{(\omega^2+{\omega^B_\pm}^2)}{4\omega\omega^B_\pm}=\dfrac{(1+\eta)}{2\sqrt{1\pm2\eta}}\simeq\frac{1}{2}+\frac{1}{4}\eta^2\\
\xi^B&=&\dfrac{(\omega-\omega_+)^2}{8\omega\omega_+}+\dfrac{(\omega-\omega_-)^2}{8\omega\omega_-}\simeq\frac{1}{4}\eta^2=\xi\;\;,
\end{array}}
\end{equation}
where the $\simeq$ equalities are valid at second order in $\eta$.
It is interesting to note that the case $\eta = 0$ (i.e., the case in the absence of light-matter interaction) cannot be immediately recovered from Eq.~(\ref{eq:effective}). In fact, in this case, from Eq.~(\ref{eq:bb}), we should have
\begin{equation}
H(\eta=0)=\omega a_1^\dagger a_1 + \omega a_2^\dagger a_2 +\omega_m b^\dagger b + g_0 a_1^\dagger a_1(b+b^\dagger)\;\;,
\end{equation}
which does not correspond to what can be found when substituting $\eta=0$ in Eq.~(\ref{eq:effective}). This is simply due to the fact that in Eq.~(\ref{eq:effective}) we neglected terms rotating at frequencies proportional to $\omega\eta$, which is justified only in the ultrastrong coupling regime.

At zero temperature, while the radiation pressure in the absence of the atom is null $P^{\eta=0}_\text{GS}$, in the presence of matter it takes a nonzero value, i.e.,  $P^{\eta}_\text{GS}={g_0}\xi/x_{zp}$ in the spin case ($g_0\xi^B/x_{zp}$ in the bosonic case). We note that, for high temperatures, the ratio between the two pressure still depends on $\eta$. While the low-energy analysis for the spin case prevent us from studying this high-temperature limit, in the bosonic case we immediately obtain
\begin{equation}
\frac{P^\eta}{P^{\eta=0}}=1+\frac{\eta^2}{2}\;\;,
\end{equation}
in the case $n_+=n_-=n$, and taking $n$ to be the occupation number for the electromagnetic environment also in the absence of matter.
\section{Interaction with environment}
\label{AppB}
In this section we show how to model the interaction with the environment for both bosonic and spin matter cases. To lighten the notation throughout the section we omit the suffix $B$ for the parameters in the bosonic case.
\subsection{Master equation for Bosonic Matter}
To correctly describe the steady-state behavior of this system we must correctly describe its interaction with three independent baths, one for each subsystem,
\begin{equation}
\label{eq:eq}
\begin{array}{lll}
H_\text{bath}&=&H^0_\text{bath}+H_\text{bath}^I\;\;,
\end{array}
\end{equation}
where
\begin{equation}
{\renewcommand{\arraystretch}{1.75}\begin{array}{lll}
H^0_\text{bath} &=&\sum_{\omega_j}\omega_j\left[t_1(\omega_j)^\dagger t_1(\omega_j)+t_2(\omega_j)^\dagger t_2(\omega_j)+t_b(\omega_j)^\dagger t_b(\omega_j)\right]\\
H_\text{bath}^I&=&\sum_{\omega_j}\left[\lambda_1(\omega_j) (a_1+a_1^\dagger)(t_1(\omega_j)+t_1^\dagger(\omega_j))+\lambda_2 (\omega_j)(a_2+a_2^\dagger)(t_2(\omega_j)+t_2^\dagger(\omega_j))+\lambda_m (\omega_j)(b+b^\dagger)(t_m+t_m^\dagger)\right]\;\;,
\end{array}}
\end{equation}
in terms of bosonic annihilation operators $t_1$, $t_2$, $t_m$ representing the baths interacting with the cavity, matter and mechanics respectively with interaction rates $\lambda_1$, $\lambda_2$, $\lambda_m$.
By using the results of the previous section, we can substitute the Bogoliubov relations in Eq.~(\ref{eq:expr}) into  $H_\text{bath}^I$ to get
\begin{equation}{\renewcommand{\arraystretch}{1.5}
\begin{array}{lll}
H_\text{bath}^I &=& \displaystyle\sum_{\omega_j}\left\{\sqrt{\dfrac{\omega}{2\omega_+}}(a_+ + a_+^\dagger)\left[\lambda_1(\omega_j)B_1(\omega_j) + \lambda_2(\omega_j) B_2(\omega_j)\right]+\sqrt{\dfrac{\omega}{2\omega_-}}(a_- + a_-^\dagger)\left[\lambda_1(\omega_j)B_1 (\omega_j)- \lambda_2(\omega_j) B_2(\omega_j)\right]\right.\\
&&\left.\left.+\lambda_m (\omega_j)(b+b^\dagger)(t_m(\omega_j)+t_m^\dagger(\omega_j))\right]\right\}\;\;,
\end{array}}
\end{equation}
where $B_1(\omega)=t_1(\omega)+t^\dagger_1(\omega)$ and $B_2(\omega)=t_2(\omega)+t_2^\dagger(\omega)$. Now, by defining
\begin{equation}
p_\pm (\omega_j)= \dfrac{\lambda_1 (\omega_j) t_1  (\omega_j)\pm \lambda_2 (\omega_j) t_2 (\omega_j)}{\sqrt{\lambda_1^2 (\omega_j)+\lambda_2^2 (\omega_j)}}\;\;,
\end{equation}
 we obtain (omitting $\omega_j$ dependences)
\begin{equation}
\begin{array}{lll}
H_\text{bath}^I &=&\sum_{\omega_j} \lambda_+(a_+ + a_+^\dagger)(p_+^\dagger + p_+)+\sum_{\omega_j} \lambda_-(a_- + a_-^\dagger)(p_-^\dagger + p_-)+\left.\sum_{\omega_j} \lambda_m (b+b^\dagger)(t_m+t_m^\dagger)\right]\;\;,
\end{array}
\end{equation}
where
\begin{equation}
\lambda_\pm=\left[\left({\frac{\omega}{\omega_j}}\right){\frac{\lambda_1^2(\omega_j)+\lambda^2_2(\omega_j)}{2}}\right]^{1/2}\;\;,
\end{equation}
and similarly for the free term
\begin{equation}
H^0_\text{bath}=\sum_{\omega_j}\omega_j\left[p_+(\omega_j)^\dagger p_+(\omega_j)+p_-(\omega_j)^\dagger p_-(\omega_j)+t_m(\omega_j)^\dagger t_m(\omega_j)\right]\;\;.
\end{equation}

This shows the normal modes interact with independent baths. In this way we can immediately integrate out the baths to obtain a master equation which can be written as \cite{Carmichael2002}
 \begin{equation}
\begin{array}{lll}
\dot{\rho}=-i[H,\rho]+\mathcal{L}_+(\rho)+\mathcal{L}_-(\rho)+\mathcal{L}_m(\rho)\;\;,
\end{array}
\end{equation}
where
\begin{equation}
\label{eq:first}{\renewcommand{\arraystretch}{1.75}
\begin{array}{lll}
\mathcal{L}_\pm(\rho)&=&\left[2\pi\sum_{\omega_j}\lambda^2_\pm(\omega_j)\langle p_\pm^\dagger p_\pm\rangle\delta(\omega_j-\omega_\pm)\right]\mathcal{D}[a_\pm^\dagger](\rho)+\left[2\pi\sum_{\omega_j}\lambda^2_\pm(\omega_j)(1+\langle p_\pm^\dagger p_\pm\rangle)\delta(\omega_j-\omega_\pm)\right]\mathcal{D}[a_\pm](\rho)\\
\mathcal{L}_m(\rho)&=&\Gamma_m \left(n(T_m)\mathcal{D}[b^\dagger](\rho)+(1+n(T_m))\mathcal{D}[b](\rho)\right)\;\;,
\end{array}}
\end{equation}
where
\begin{equation}{\renewcommand{\arraystretch}{1.75}
\begin{array}{lll}
\mathcal{D}[O](\rho)&=&\frac{1}{2}(2O\rho O^\dagger-\rho O^\dagger O-O^\dagger O \rho)\\
\Gamma_m&=&2\pi d_m \lambda_m^2\;\;,
\end{array}}
\end{equation}
 where $d_m$ is the density of states for the mechanical bath associated with temperature $T_m$. By introducing the densities of states $d_1$, $d_2$ for the remaining modes, we can further simplify this result by explicitly computing
\begin{equation}
\label{eq:second}{\renewcommand{\arraystretch}{1.75}
\begin{array}{lll}
\sum_{\omega_j}\lambda^2_\pm(\omega_j)(1+\langle p_\pm^\dagger p_\pm\rangle)\delta(\omega_j-\omega_\pm)&=&\sum_{\omega_j}\dfrac{\omega}{\omega_\pm}\dfrac{\lambda_1^2+\lambda_2^2}{2}\left(1+\dfrac{\lambda_1^2 n(\omega_j,T_1)+\lambda_2^2 n(\omega_j,T_2)}{\lambda_1^2+\lambda_2^2}\right)\delta(\omega_j-\omega_\pm)\\
&=&\dfrac{\omega}{2\omega_\pm}\int d\bar{\omega}~\delta(\bar{\omega}-\omega_\pm)\left[\lambda_1^2(\bar{\omega}) d_1(\bar{\omega})(1+\bar{n}(\bar{\omega},T_1))+\lambda_2^2(\bar{\omega}) d_2(\bar{\omega})(1+\bar{n}(\bar{\omega},T_1)) \right]\\
&=&\dfrac{\omega}{2\omega_\pm}\left[\lambda_1^2(\omega_\pm) d_1(\omega_\pm)(1+\bar{n}(\omega_\pm,T_1))+\lambda_2^2(\omega_\pm) d_2(\omega_\pm)(1+\bar{n}(\omega_\pm,T_1)) \right]\\
\sum_{\omega_j}\lambda^2_\pm(\omega_j)\langle p_\pm^\dagger p_\pm\rangle \delta(\omega_j-\omega_\pm)&=&\dfrac{\omega}{2\omega_\pm}\left[\lambda_1^2(\omega_\pm) d_1(\omega_\pm)\bar{n}(\omega_\pm,T_1)+\lambda_2^2(\omega_\pm) d_2(\omega_\pm)\bar{n}(\omega_\pm,T_1) \right]\;\;.
\end{array}}
\end{equation}

We  now define
\begin{equation}{\renewcommand{\arraystretch}{1.75}
\begin{array}{lll}
\kappa_\pm&=&2\pi\dfrac{\omega}{2\omega_\pm}\left[d_1(\omega_\pm)\lambda^2_1(\omega_\pm)+d_2(\omega_\pm)\lambda^2_2(\omega_\pm)\right]\\
\kappa_\pm n_\pm&=&2\pi\dfrac{\omega}{2\omega_+}\left[d_1(\omega_\pm)\lambda^2_1(\omega_\pm)\bar{n}(\omega_\pm,T_1)+d_2(\omega_\pm)\lambda^2_2(\omega_\pm)\bar{n}(\omega_\pm,T_2)\right]\;\;,
\end{array}}
\end{equation}
where $\bar{n}_\pm=\bar{n}(\omega_\pm,T_\pm)$ and where we introduced the effective temperatures $T_\pm$ which can be found by solving the implicit equation
\begin{equation}
 \bar{n}_\pm=\frac{d_1(\omega_\pm)\lambda^2_1(\omega_\pm)\bar{n}(\omega_\pm,T_1)+d_2(\omega_\pm)\lambda^2_2(\omega_\pm)\bar{n}(\omega_\pm,T_2)}{d_1(\omega_\pm)\lambda^2_1(\omega_\pm)+d_2(\omega_\pm)\lambda^2_2(\omega_\pm)}\;\;,
\end{equation}
as
\begin{equation}
k_{\mathrm{B}}T_\pm={\hbar\omega_\pm}{\left[\log{\left(1+\frac{1}{\bar{n}(\omega_\pm,T_1)+\bar{n}(\omega_\pm,T_2)}\right)}\right]^{-1}}\;\;.
\end{equation}

 As an immediate check, if we assume $T_1=T_2=T$ we get $\bar{n}_\pm=\bar{n}(\omega_\pm,T)$.
In this way we can write the following simplified form for the Liouvillians
\begin{equation}
\begin{array}{lll}
\mathcal{L}_\pm(\rho)&=& \kappa_\pm\left(\bar{n}_\pm\mathcal{D}[a_\pm^\dagger](\rho)+(1+\bar{n}_\pm)\mathcal{D}[a_\pm](\rho)\right)\;\;,
\end{array}
\end{equation}
where
\begin{equation}
\kappa_\pm = \frac{\omega}{\omega_\pm}\frac{\kappa_1+\kappa_2}{2}\;\;,
\end{equation}
with
\begin{equation}{\renewcommand{\arraystretch}{1.75}
\begin{array}{lll}
\kappa_1&=&2\pi d_1(\omega_\pm)\lambda_1^2(\omega_\pm)\\
\kappa_2&=&2\pi d_2(\omega_\pm)\lambda_2^2(\omega_\pm)\;\;,
\end{array}}
\end{equation}
are the rates if the systems were independently coupled to their baths (but evaluated at the polaritonic frequencies).

This result is general and has no temperature restrictions in this bosonic matter case. Below, we derive a master equation which can be used to model both the spin and bosonic matter cases in the low energy limit. When applied to the bosonic case, the result will match the one given in this section, as it logically should.

\subsection{Master equation at low temperatures (for both boson and spin cases)}
As mentioned above, in the ultrastrong coupling regime, the environment induces transitions between the dressed eigenstates. The master equation for the system can be written as \cite{Beaudoin2011}
\begin{equation}
\label{eq:lin22}
\begin{array}{lll}
\dot{\rho}=-i[H,\rho]+\mathcal{L}_1(\rho)+\mathcal{L}_2(\rho)+\mathcal{L}_m(\rho)\;\;,
\end{array}
\end{equation}
where
\begin{equation}{\renewcommand{\arraystretch}{1.75}
\begin{array}{lll}
\mathcal{L}_{1}(\rho)&=& \sum_{j,k>j}\Gamma^{jk}_1\bar{n}(\Delta_{kj},T_1)\mathcal{D}[\ket{k}\bra{j}](\rho)+\sum_{j,k>j}\Gamma^{jk}_1(1+\bar{n}(\Delta_{kj},T_1))\mathcal{D}[\ket{j}\bra{k}](\rho)\\
\mathcal{L}_{2}(\rho)&=& \sum_{j,k>j}\Gamma^{jk}_2\bar{n}(\Delta_{kj},T_2)\mathcal{D}[\ket{k}\bra{j}](\rho)+\sum_{j,k>j}\Gamma^{jk}_2(1+\bar{n}(\Delta_{kj},T_2))\mathcal{D}[\ket{j}\bra{k}](\rho)\\
\mathcal{L}_m(\rho)&=&\kappa\left(n(T_m)\mathcal{D}[b^\dagger](\rho)+(1+n(T_m))\mathcal{D}[b](\rho)\right)\;\;,
\end{array}}
\end{equation}
where $H=H^\text{eff},H^B$, and where the indexes $j,k$ label eigenstates of the system in increasing energy order and where
\begin{equation}
\begin{array}{lll}
\Gamma^{jk}_1&=&2\pi d_1(\Delta_{kj})\lambda^2_1(\Delta_{jk})|\bra{k}(a_1+a_1^\dagger)\ket{j}|^2\\
\Gamma^{jk}_2&=&2\pi d_2(\Delta_{kj})\lambda^2_2(\Delta_{jk})|\bra{k}(a_2+a_2^\dagger)\ket{j}|^2\;\;,
\end{array}
\end{equation}
in terms of the density of states of the bath $d_1$ and $d_2$ in the bosonic case and
\begin{equation}
\begin{array}{lll}
\Gamma^{jk}_1&=&2\pi d_1(\Delta_{kj})\lambda^2_1(\Delta_{jk})|\bra{k}(a_1+a_1^\dagger)\ket{j}|^2\\
\Gamma^{jk}_2&=&2\pi d_2(\Delta_{kj})\lambda^2_2(\Delta_{jk})|\bra{k}(\sigma_-+\sigma_+)\ket{j}|^2\;\;,
\end{array}
\end{equation}
 in the spin case (in this section the cavity mode will be denoted with $a_1$). We explicitly note that, as explained in  \cite{Beaudoin2011}, the degeneracies present in the bosonic case should pose a problem in the derivation of Eq.~(\ref{eq:lin22}). However, in this case, the degeneracies are lifted at an effective level, by imposing the low energy approximation.
This energy restriction amounts to considering as the only relevant states for the fields $a_1$ and $a_2$  the ground and first excited states. We can then write, for example
\begin{equation}
\label{eq:se}
\begin{array}{lll}
\sum_{j,k>j}\Gamma^{jk}_1\bar{n}(\Delta_{kj},T_1)\mathcal{D}[\ket{k}\bra{j}](\rho)&\simeq& 2\pi d_1(\omega_+)\lambda^2_1(\omega_+)|\bra{+} (a_1+a_1^\dagger)\ket{G}|^2\bar{n}(\omega_+,T_1)\mathcal{D}[\ket{+}\bra{G}](\rho)\\
&&+2\pi d_1(\omega_-)\lambda^2_1(\omega_-)|\bra{-} (a_1+a_1^\dagger)\ket{G}|^2\bar{n}(\omega_-,T_1)\mathcal{D}[\ket{-}\bra{G}](\rho)\;\;.
\end{array}
\end{equation}

This is the point where differences due to the the spin or bosonic nature of the matter degree of freedom enter the analysis.  This is simply due to different expressions for the transition matrix elements.
\subsubsection{Bosonic case}
In the bosonic case, we can use the Bogoliubov transformations in Eq.~(\ref{eq:expr}) and evaluate the previous expression as
\begin{equation}
\begin{array}{lll}
\displaystyle\sum_{j,k>j}\Gamma^{jk}_1\bar{n}(\Delta_{kj},T_1)\mathcal{D}[\ketbra{k}{j}](\rho)&\simeq& 2\pi[\rho_1(\omega_+)\lambda^2_1(\omega_+)\dfrac{\omega}{2\omega_+}\bar{n}(\omega_+,T_1)\mathcal{D}[a^\dagger_+](\rho)+ \rho_1(\omega_-)\lambda^2_1(\omega_-)\dfrac{\omega}{2\omega_-}\bar{n}(\omega_-,T_1)\mathcal{D}[a^\dagger_-](\rho)]\;\;,
\end{array}
\end{equation}
and analogously for the remaing terms in Eq.~(\ref{eq:lin22}) to get
 \begin{equation}
\begin{array}{lll}
\dot{\rho}=-i[H,\rho]+\mathcal{L}_+(\rho)+\mathcal{L}_-(\rho)+\mathcal{L}_m(\rho)\;\;,
\end{array}
\end{equation}
where
\begin{equation}{\renewcommand{\arraystretch}{1.75}
\begin{array}{lll}
\mathcal{L}_\pm(\rho)&=& 2\pi\dfrac{\omega}{2\omega_\pm}\left[d_1(\omega_+)\lambda^2_1(\omega_\pm)\bar{n}(\omega_\pm,T_1)+d_2(\omega_\pm)\lambda^2_2(\omega_\pm)\bar{n}(\omega_\pm,T_2)\right]\mathcal{D}[a_\pm^\dagger](\rho)\\
&&+2\pi\dfrac{\omega}{2\omega_\pm}\left[d_1(\omega_\pm)\lambda^2_1(\omega_\pm)(1+\bar{n}(\omega_\pm,T_1))+d_2(\omega_\pm)\lambda^2_2(\omega_\pm)(1+\bar{n}(\omega_\pm,T_2))\right]\mathcal{D}[a_\pm](\rho)\;\;,
\end{array}}
\end{equation}
which, as promised, is in fact equivalent to the expression given in Eq.~(\ref{eq:first}) (by immediate use of Eq.~(\ref{eq:second})), and all the subsequent analysis can be taken from there.
\subsubsection{Spin case}
In the spin case, the matrix coefficients take a different form which can be computed at second order in $\eta$ by Eq.~(\ref{eq:fstates}), and explicitly read
\begin{equation}{\renewcommand{\arraystretch}{1.75}
\begin{array}{lllll}
\zeta^\pm_a\equiv|\bra{\pm}(a+a^\dagger)\ket{G}|^2&=&\frac{1}{2}|(1\mp\frac{3\eta}{4}+\frac{15\eta^2}{32})|^2&=&\frac{1}{2}(1\mp\frac{\eta}{4}+\frac{\eta^2}{32})\\
\zeta^\pm_\sigma\equiv|\bra{\pm}(\sigma_-+\sigma_+)\ket{G}|^2&=&\frac{1}{2}|(1\mp\frac{\eta}{4}+\frac{\eta^2}{32})|^2&=&\frac{1}{2}(1\mp\frac{\eta}{2}+\frac{\eta^2}{8})\;\;.
\end{array}}
\end{equation}

 Consequently, Eq.~(\ref{eq:se}) in this case becomes
\begin{equation}
\begin{array}{lll}
\displaystyle\sum_{j,k>j}\Gamma^{jk}_1\bar{n}(\Delta_{kj},T_1)\mathcal{D}[\ketbra{k}{j}](\rho)&\simeq& 2\pi[\rho_1(\omega_+)\lambda^2_1\zeta^+_a\bar{n}(\omega_+,T_1)\mathcal{D}[a^\dagger_+](\rho)+ \rho_1(\omega_-)\lambda^2_1(\omega_-)\zeta^-_a\bar{n}(\omega_-,T_1)\mathcal{D}[a^\dagger_-](\rho)]\\
\displaystyle\sum_{j,k>j}\Gamma^{jk}_2\bar{n}(\Delta_{kj},T_2)\mathcal{D}[\ketbra{k}{j}](\rho)&\simeq& 2\pi[\rho_2(\omega_+)\lambda^2_2\zeta^+_\sigma\bar{n}(\omega_+,T_2)\mathcal{D}[a^\dagger_+](\rho)+ \rho_2(\omega_-)\lambda^2_2(\omega_-)\zeta^-_\sigma\bar{n}(\omega_-,T_2)\mathcal{D}[a^\dagger_-](\rho)]\;\;,
\end{array}
\end{equation}
leading to formally the same solution as in the bosonic case but with different rates
\begin{equation}{\renewcommand{\arraystretch}{1.75}
\begin{array}{lll}
\kappa_\pm&=&2\pi\left[d_1(\omega_\pm)\zeta_a^\pm\lambda^2_1(\omega_\pm)+d_2(\omega_\pm)\zeta_\sigma^\pm\lambda^2_2(\omega_\pm)\right]\\
\kappa_\pm n_\pm&=&2\pi\left[d_1(\omega_\pm)\zeta_a^\pm\lambda^2_1(\omega_\pm)\bar{n}(\omega_\pm,T_1)+d_2(\omega_\pm)\zeta_\sigma^\pm\lambda^2_2(\omega_\pm)\bar{n}(\omega_\pm,T_2)\right]\;\;.
\end{array}}
\end{equation}

In the case $T_1=T_2=T$ we get $\bar{n}_\pm=\bar{n}(\omega_\pm,T)$ as in the previous case and we can write the following simplified form for the Liouvillians
\begin{equation}
\begin{array}{lll}
\mathcal{L}_\pm(\rho)&=& \kappa_\pm\left(\bar{n}_\pm\mathcal{D}[a_\pm^\dagger](\rho)+(1+\bar{n}_\pm)\mathcal{D}[a_\pm](\rho)\right)\;\;,
\end{array}
\end{equation}
where $\kappa_\pm = \zeta_a^\pm\kappa_1+\zeta_\sigma^\pm\kappa_2$ with $\kappa_1=2\pi d_1(\omega_\pm)\lambda_1^2(\omega_\pm)$ and $\kappa_2=2\pi d_1(\omega_\pm)\lambda_1^2(\omega_\pm)$ are the rates if the systems were independently coupled to their baths (but evaluated at the polaritonic frequencies).
\section{Heisenberg equation of motion}
\label{AppC}
In this section we first derive an expression for the quadrature averages and variances for the spin and bosonic cases. The different physical nature of these two models enters the derivation through a different expression for the parameters, as summarized in the following table.
{\renewcommand{\arraystretch}{2}
\begin{center}
    \begin{tabularx}{\textwidth}{ |c|Y|Y|Y|Y|}
    \hline
     & $\omega_\pm$&${\mathbf{\alpha_\pm}}$& $\xi$&$\kappa_\pm$\\ \hline
             \textbf{Spin Case}&$(1\pm\eta)\omega$&$\frac{1}{2}\mp\frac{\eta}{4}$&$\frac{\eta^2}{4}$&$ \zeta_a^\pm\kappa_1+\zeta_\sigma^\pm\kappa_2$\\\hline
   \textbf{Bosonic Case} & $\omega({1\pm2\eta})^{1/2}$&$\frac{(\omega^2+{\omega_\pm}^2)}{4\omega\omega_\pm}\simeq\frac{1}{2}+\frac{1}{4}\eta^2$&$\frac{(\omega-\omega_+)^2}{8\omega\omega_+}+\frac{(\omega-\omega_-)^2}{8\omega\omega_-}\simeq\frac{\eta^2}{4}$&$ \frac{\omega}{\omega_\pm}\frac{\kappa_1+\kappa_2}{2}$\\
    \hline
    \end{tabularx}
\end{center}}
For the spin case, the validity of the model is restricted to a low energy limit and at second order in $\eta$, which we require to be $\eta\simeq0.1$ (ultrastrong coupling regime) to derive the master equation. The bosonic model is, in principle, exact at all temperatures. However, we notice that the rotating-wave approximation applied to obtain Eq.~(\ref{eq:effective}) also requires the condition $\eta^3\gg{g_0}/{\omega}$ to be satisfied.

Finally, in the last subsection, we then use the expressions for the quadratures to quantify the visibility of the effect considered in this article, i.e., the regime where virtual radiation pressure is observable for both bosonic and spin cases.
\subsection{Solution}
Let us now consider the Hamiltonian in Eq.~(\ref{eq:effective})
\begin{equation}
H= \omega_+a^\dagger_+a_++\omega_-a^\dagger_-a_-+\omega_m b^\dagger b +\sqrt{2} g_0 {X} \hat{\alpha}\;\;,
\end{equation}
where the dimensionless quadratures are defined as ${X} =({b+b^\dagger})/{\sqrt{2}}$ and ${\tilde{X}} =i({b^\dagger-b})/{\sqrt{2}}$, and where $\hat{\alpha} = \alpha_+ a_+^\dagger a_+ + \alpha_- a_-^\dagger a_- + \xi$.

The state of the system is described by a density matrix which satisfies
\begin{equation}
\begin{array}{lll}
\dot{\rho}=-i[H,\rho]+\mathcal{L}_+(\rho)+\mathcal{L}_-(\rho)+\mathcal{L}_m(\rho)\;\;,
\end{array}
\end{equation}
where $\mathcal{L}_\pm(\rho)=\kappa_\pm\left(\bar{n}_\pm\mathcal{D}[a_\pm^\dagger](\rho)+(1+\bar{n}_\pm)\mathcal{D}[a_\pm](\rho)\right)$, and $\mathcal{L}_m(\rho)=\Gamma_m\left(\bar{n}_m\mathcal{D}[b^\dagger](\rho)+(1+\bar{n}_m)\mathcal{D}[b](\rho)\right)$, and where $\bar{n}_\pm=n(T_\pm)$ and $\bar{n}_m = n(T_m)$.
The Heisenberg equation of motion for a generic operator $O$ can be written as
\begin{equation}
{\renewcommand{\arraystretch}{1.75}
\begin{array}{lll}
\langle\dot{O}\rangle&=&-i\langle[O,H]\rangle\\
&&+\frac{\kappa_+}{2}\bar{n}_+\left(\langle[[a_+,O],a^\dagger_+]\rangle+\langle[[a_+^\dagger,O],a_+]\rangle\right)+\frac{\kappa_+}{2}\left(\langle[a_+^\dagger,O]a_+\rangle+\langle a^\dagger_+[O,a_+]\rangle\right)\\
&&+\frac{\kappa_-}{2}\bar{n}_-\left(\langle[[a_-,O],a^\dagger_-]\rangle+\langle[[a_-^\dagger,O],a_-]\rangle\right)+\frac{\kappa_-}{2}\left(\langle[a_-^\dagger,O]a_-\rangle+\langle a^\dagger_-[O,a_-]\rangle\right)\\
&&+\frac{\Gamma_m}{2}\bar{n}_m\left(\langle[[b,O],b^\dagger]\rangle+\langle[[b^\dagger,O],b]\rangle\right)+\frac{\Gamma_m}{2}\left(\langle[b^\dagger,O]b\rangle+\langle b^\dagger[O,b]\rangle\right)\;\;.
\end{array}}
\end{equation}

In the steady state
\begin{equation}
\label{eq:averages}
\begin{array}{lll}{\renewcommand{\arraystretch}{1.75}
\begin{array}{cll}
\langle a^\dagger_\pm a_\pm\rangle&=&\bar{n}_\pm\\
\langle (a^\dagger_\pm a_\pm)^2\rangle&=&\bar{n}_\pm+2\bar{n}_\pm^2\\
\langle a^\dagger_+ a_+ a^\dagger_- a_-\rangle&=&\bar{n}_+\bar{n}_-
\end{array}}&~~~~~~~~~~~&
{\renewcommand{\arraystretch}{1.75}\begin{array}{lll}
\langle {X} \rangle&=&-4 \sqrt{2}\alpha \bar{\eta}_m\dfrac{ Q_m}{1+4 Q_m^2}\\
\langle {\tilde{X}}  \rangle &=&-2\sqrt{2}\alpha\bar{\eta}_m\dfrac{1}{1+4 Q_m^2}\;\;,
\end{array}}
\end{array}
\end{equation}
where
\begin{equation}{\renewcommand{\arraystretch}{1.75}
\begin{array}{lll}
Q_m &=& \dfrac{\omega_m}{\Gamma_m}\\
\bar{\eta}_m& =& \dfrac{g_0}{\Gamma_m}=\eta_m Q_m\;\;,
\end{array}}
\end{equation}
 and
\begin{equation}
\label{eq:alfa}
\alpha=\langle\hat{\alpha}\rangle=\alpha_+\bar{n}_++\alpha_-\bar{n}_-+\xi\;\;.
\end{equation}
The cavity zero-point energy contributions neglected in Eq.~(\ref{eq:averages}), would lead to an additional temperature-independent term in the previous  expression which can be obtained by the replacement $\alpha\mapsto\alpha+1/2$. Similarly, we can calculate the correlations between the light-matter system and the mechanical mode as
\begin{equation}{\renewcommand{\arraystretch}{1.75}
\begin{array}{lll}
\langle a^\dagger_\pm a_\pm {X} \rangle&=&\langle {X} \rangle \bar{n}_\pm-p_\pm\\
\langle a^\dagger_\pm a_\pm {\tilde{X}} \rangle&=&\bar{n}_\pm\langle {\tilde{X}} \rangle-s_\pm\;\;,
\end{array}}
\end{equation}
where
\begin{equation}{\renewcommand{\arraystretch}{2}
\begin{array}{lll}
p_\pm&=&4\sqrt{2}\dfrac{ Q_m \bar{\eta}_m \alpha_\pm \bar{n}_\pm(1+\bar{n}_\pm)}{4 Q_m^2+{\beta_\pm}^2}\\ {\beta_\pm}&=&\dfrac{\Gamma_m+2\kappa_{\pm}}{\Gamma_m}\\
s_\pm&=&2\sqrt{2}\dfrac{{\beta_\pm} \bar{\eta}_m \alpha_\pm \bar{n}_\pm(1+\bar{n}_\pm)}{4 Q_m^2+{\beta_\pm}^2}\;\;.
\end{array}}
\end{equation}

The mechanical correlations are readily found to be
\begin{equation}\label{eq:threeCorr}\left\{{\renewcommand{\arraystretch}{2}
\begin{array}{lll}
\langle {X} {\tilde{X}} +{\tilde{X}} {X} \rangle&=&\dfrac{2\sqrt{2}}{1+4 Q_m^2}(-2\alpha \langle X\rangle+p+2 Q_m s)\\
\langle {X} ^2\rangle&=&\bar{n}_m+\frac{1}{2}+\langle {X} \rangle^2+2\sqrt{2}\dfrac{\bar{\eta}_m Q_m}{1+4 Q_m^2}(p+2 Q_m s)\\
\langle {\tilde{X}} ^2\rangle &=& \bar{n}_m+\frac{1}{2}+\langle {\tilde{X}} \rangle^2-2\sqrt{2}\dfrac{\bar{\eta}_m Q_m}{1+4 Q_m^2}p+2\sqrt{2}\dfrac{\bar{\eta}_m}{1+4 Q_m^2}(1+2 Q_m^2)s\;\;,
\end{array}}\right.
\end{equation}
where $p=\alpha_+p_++\alpha_-p_-$ and $s=\alpha_+s_++\alpha_-s_-$, leading to
\begin{equation}
\label{eq:deltaX}
\begin{array}{lll}
\delta{\tilde{X}} ^2&=&\langle {\tilde{X}} ^2\rangle-\langle {\tilde{X}} \rangle^2\\
&\leq&\frac{1}{2}+\bar{n}_m+2\sqrt{2}\dfrac{\bar{\eta}_m}{1+4 Q_m^2}(1+2 Q_m^2)s\\
&\leq&\frac{1}{2}+\bar{n}_m+8\bar{\eta}_m^2 R\;\;,
\end{array}
\end{equation}
where we used the fact that $\text{argmax}[f( Q_m)]=0$, where
\begin{equation}
f(Q_m)=\frac{(1+2 Q_m^2)}{(1+4 Q_m^2)(Q^2+4 Q_m^2)}\;\;,
\end{equation}
 and  defined
\begin{equation}
\label{eq:R}
R=\frac{\alpha^2_+}{{\beta_+}}\bar{n}_+ (1+\bar{n}_+)+\frac{\alpha^2_-}{{\beta_-}}\bar{n}_-(1+\bar{n}_-)\;\;.
\end{equation}
Note that the quantity $\langle\tilde{X}^2\rangle$ is affected from the cavity zero-point energy contributions only through $\langle\tilde{X}\rangle^2$ in Eq.~(\ref{eq:threeCorr}). For this reason, the expression for the variance $\delta\tilde{X}^2$ is independent from such contributions.\\

For completeness, we also report the results for the variance of the other quadrature
\begin{equation}
\begin{array}{lll}
\delta\!{X} ^2&=&\langle {X} ^2\rangle-\langle {X} \rangle^2\\
&\leq&\bar{n}_m+\frac{1}{2}+16 \bar{\eta}_m^2 \tilde{R}\;\;,
\end{array}
\end{equation}
 where we used
 \begin{equation}
 \text{argmax}(f_{X} ( Q_m))=\frac{\sqrt{Q}}{2}\;\;,
 \end{equation}
  and defined
  \begin{equation}
  \tilde{R}=\frac{\alpha_+^2 \bar{n}_+ (1+\bar{n}_+)}{4(1+{\beta_+})}+\frac{\alpha_-^2 \bar{n}_- (1+\bar{n}_-)}{4(1+{\beta_-})}\;\;.
  \end{equation}
\subsection{Visibility}
The modulation of the opto-mechanical coupling $g_0\mapsto g_0\cos{(\omega_d t)}$  effectively corresponds (in a frame rotating at $\omega_d$) to the redefinitions
\begin{equation}\label{eq:redef}{\renewcommand{\arraystretch}{1.85}
\begin{array}{clc}
g_0&\mapsto& {g_0}/{2}\\
\bar{\eta}_m&\mapsto&{\bar{\eta}_m}/{2}\\
\omega_m&\mapsto&\delta\;\;,
\end{array}}
\end{equation}
where, as a reminder, $\bar{\eta}_m=g_0/\Gamma_m$ and $\delta=\omega_m-\omega_d\ll\omega_m$.
In this way, from Eq.~(\ref{eq:averages}), the displacement of the quadrature $\tilde{X}$ can be written as
\begin{equation}
\label{eq:aa}
|\langle {\tilde{X}}  \rangle |=|\langle {\tilde{X}}  \rangle_{\bar{n}} |+|\langle {\tilde{X}}  \rangle_\text{GS} |\;\;,
\end{equation}
where
\begin{equation}
\begin{array}{lll}
|\langle {\tilde{X}} \rangle_{\bar{n}}|&=&\dfrac{\sqrt{2}\bar{\eta}_m}{1+4Q^2_\delta}(\alpha-\xi)\\
|\langle {\tilde{X}}  \rangle_\text{GS}|&=&\dfrac{\sqrt{2}\bar{\eta}_m}{1+4Q_\delta^2}\xi\;\;,
\end{array}
\end{equation}
are the thermal (associated with the index $\bar{n}$) and ground state (associated with the index $\text{GS}$) contributions to the total displacement respectively  (with $Q_\delta={\delta}/{\Gamma_m}$). Note that these equations differ from Eqs.~(\ref{eq:averages}) by a factor 2 due to the re-definitions outlined in Eq.~(\ref{eq:redef}). Moreover, the zero-point contribution to the cavity energy calculated in the previous section would effectively add a constant term $|\langle \tilde{X}\rangle|_{zp}={\bar{\eta}_m}/{\sqrt{2}(1+4Q_\delta^2)}$ to the expression in Eq.~(\ref{eq:aa}). However, as seen from our previous analysis, such a contribution does not affect the expression for the variances. For this reason, it can simply be subtracted off the average value.\\

The signal we are interested to resolve is the displacement due to ground-state effects, i.e., $\langle\tilde{X}\rangle_\text{GS}$. This  makes it natural to define signal-to-noise ratio
in the following way
\begin{equation}
\label{eq:FGS}
F_\text{GS}=\dfrac{|\langle\tilde{X}\rangle_\text{GS}|}{\delta\!\tilde{X}}\;\;,
\end{equation}
where, from Eq.~(\ref{eq:deltaX})
\begin{equation}
\delta{\tilde{X}} ^2\leq\frac{1}{2}+\bar{n}_m+2\bar{\eta}_m^2 R\;\;,
\end{equation}
where the missing factor $4$ in front of $\bar{\eta}_m$ takes into account Eq.~(\ref{eq:redef}).

We then want to impose two conditions for the observation of the effect. The first, is the standard quantum limit (SQL) requirement
\begin{equation}
F_\text{GS}>1\;\;,
\end{equation}
for the resolution of the signal. Secondly, we require to be in a regime where  ground state effects are predominant with respect to thermal ones, i.e.,
\begin{equation}
|\langle\tilde{X}\rangle_{\text{GS}}|>|\langle \tilde{X}\rangle_{\bar{n}}|\;\;.
\end{equation}
Alternatively, dividing both sides by $\delta\!\tilde{X}$, this condition can be written as $F_{\text{GS}}>F_{\bar{n}}$, where $F_{\bar{n}}=F-F_{\text{GS}}$. In this way, we are equivalently requiring that most of the resolved physical ratio
\begin{equation}
F=\dfrac{\langle\tilde{X}\rangle}{\delta\!\tilde{X}}
\end{equation}
 is due to ground state effects. Let us now analyze both of these in more detail.
From Eq.~(\ref{eq:aa}) we see that the condition for the predominance of ground state effects $F_\text{GS}>1$ takes the simple form $\xi>\alpha-\xi$ or
\begin{equation}
\label{eq:nGS}
\bar{n}<n_\text{GS}=\frac{\eta^2}{4}\;\;,
\end{equation}
where we used the definition in Eq.~(\ref{eq:alfa}), and the expressions for $\alpha_\pm$ and $\xi$ for the spin and bosonic case. Up to second-order in $\eta$ this expression leads to the result in Eq.~(\ref{eq:nGS}).

The SQL condition $F_\text{GS}>1$ is more complex and involves the mechanical variance. Using the definition in Eq.~(\ref{eq:FGS}), together with the results in Eqs. (\ref{eq:averages}) and(\ref{eq:deltaX}) we obtain, through some algebra
\begin{equation}
\label{eq:gg}
\bar{\eta}_m>\bar{\eta}^\text{SQL}_m\;\;,
\end{equation}
where
\begin{equation}
(\bar{\eta}^{\text{SQL}}_m)^2=\frac{(1+4 Q_\delta^2)^2(1+2\bar{n}_m)}{4(\xi^2-(1+4 Q_\delta^2)^2 R)}\;\;.
\end{equation}

The positivity of the left-hand side requires that $R<\frac{\xi^2}{(1+4 Q_\delta^2)^2}$, which, to forth-order in $\eta$, leads to
\begin{equation}
\label{eq:sql}
\bar{n}<n_\text{SQL}\;\;,
\end{equation}
where
\begin{equation}
n_\text{SQL}=\frac{1}{8}\beta\frac{\eta^4}{(1+4 Q_\delta^2)^2}\;\;.
\end{equation}
We note that the extrapolation of this result at fourth order in $\eta$ is valid at fourth-order perturbation theory since it depends quadratically on the ground state displacement whose lowest order expansion in $\eta$ is $O(\eta^2)$), and
\begin{equation}
\beta = \frac{\Gamma_m+2\gamma}{\Gamma_m}\;\;,
\end{equation}
 where $\gamma=\gamma_\pm$ in the case $\eta=0$. Note that the bound in Eq.~(\ref{eq:gg}) explicitly depends on the expressions for $\alpha_\pm$ through the quantity $R$ (defined in Eq.~(\ref{eq:R})). For this reason its expression as a function of $\eta$ will depend upon the spin or bosonic case considered. However, the bound in Eq.~(\ref{eq:sql}) is, at lowest order in $\eta$, common to the two cases.

 We can then collect these two conditions to find that
\begin{equation}
\bar{n}<n_\text{max}\;\;,
\end{equation}
where
\begin{equation}
n_\text{max}=\text{min}\left(\frac{\eta^2}{4},\frac{1}{8}\beta\frac{\eta^4}{(1+4 Q_\delta^2)^2}\right)\;\;,
\end{equation}
 is the maximum allowed occupation number for the light-matter system in order to observe the effect. More precisely, we can say that $n_\text{max}$ is the maximum occupation number such that a value of $g_0$ exists for which the effect can be observed. Such a value is given by Eq.~(\ref{eq:gg}), which,  for $ Q_\delta\rightarrow 0$ and $\bar{n}=\bar{n}_m=0$ gives
\begin{equation}
\bar{\eta}_m>\frac{2}{\eta^2}\;\;,
\end{equation}
which sets the best-conditions limit for the observation of the effect.\\
We summarize the logic and findings of this chapter in the following table.\\
{\renewcommand{\arraystretch}{2}
\begin{center}
    \begin{tabularx}{\linewidth}{ | Y| Y | Y|}
    \hline
    \textbf{Requirement} &\textbf{Condition}& \textbf{Physical Contraints}  \\ \hline
    Standard Quantum Limit & $F_{\text{GS}}>1$&$\left.\begin{array}{lll}\bar{n}&<&n_{\text{SQL}}\\\bar{\eta}_m&>&\eta^\text{SQL}_m\end{array}\right.$\\ \hline
        Ground State Effects Physics& $|\langle\tilde{X}\rangle_{\text{GS}}|>|\langle\tilde{X}\rangle_{\bar{n}}|$&$\bar{n}<n_{\text{GS}}$\\
    \hline
    \end{tabularx}
\end{center}}
For bookkeeping, the other quadrature gives the following result
\begin{equation}
\begin{array}{lll}
F_{X} ^2&=&\dfrac{\langle {X} \rangle^2}{\delta\!{X}^2}\\
&\geq&\dfrac{8\alpha^2\bar{\eta}_m^2 Q_\delta^2}{(1+4 Q_\delta^2)^2 (\frac{1}{2}+\bar{n}_m+4 \bar{\eta}^2_m\tilde{R})}\;\;,
\end{array}
\end{equation}
which, with respect to $F_{\tilde{X}}$, is suppressed by a factor $ Q_\delta$, which tends towards zero in the amplification approach considered in this article.\\
Finally, we note that, in the limit for $n_\pm=n\rightarrow\infty$, and for $\delta=0$ (i.e., driving of the opto-mechanical coupling in resonance to the mechanical frequency) we obtain
\begin{equation}
F\rightarrow2\left(\frac{1}{\beta_+}+\frac{1}{\beta_-}\right)^{-\frac{1}{2}}\;\;.
\end{equation}
In the same limit, in the absence of matter, we find
\begin{equation}
F_{\eta=0}\rightarrow\sqrt{\beta}\;\;.
\end{equation}
where $\beta = \Gamma_m/(\Gamma_m+2\kappa_1)$.

\bibliography{Ref}